\documentclass[aps,prl,preprint,superscriptaddress,showpacs,amssymb]{revtex4}
\usepackage{bm}
\usepackage{epsfig}
\begin{document}
%
%
\title{Spin-Hall effect and spin-coherent excitations in a strongly
confined two-dimensional hole gas}
%
%
\author{P. Kleinert}
\affiliation{Paul-Drude-Intitut f\"ur Festk\"orperelektronik,
Hausvogteiplatz 5-7, 10117 Berlin, Germany}
\author{V.V. Bryksin}
\affiliation{Physical Technical Institute, Politekhnicheskaya 26,
194021 St. Petersburg, Russia}
\date{\today}
\begin{abstract}
Based on a rigorous quantum-kinetic approach, spin-charge coupled
drift-diffusion equations are derived for a strongly confined
two-dimensional hole gas. An electric field leads to a coupling
between the spin and charge degrees of freedom. For weak
spin-orbit interaction, this coupling gives rise to the intrinsic
spin-Hall effect. There exists a threshold value of the spin-orbit
coupling constant that separates spin diffusion from ballistic
spin transport. In the latter regime, undamped spin-coherent
oscillations are observed. This result is confirmed by an exact
microscopic approach valid in the ballistic regime.
\end{abstract}



\maketitle


The generation and manipulation of a spin polarization by
exclusively electronic means in nonmagnetic semiconductors at room
temperature is a major challenge of spintronics. Among many
interesting phenomena, the intrinsic spin-Hall effect (SHE)
\cite{SCIE_1348,PRL_126603} has recently attracted considerable
interest. Experimental studies
\cite{SCIE_1910,PRL_047204,PRL_096605} reveal an electric-field
induced spin accumulation near the edges of a confined
two-dimensional electron (hole) gas. Most theoretical
interpretations of these experimental data rely on the notion of a
spin current oriented transverse to the applied electric
field.~\cite{SCIE_1348,PRL_126603} Interestingly, this seemingly
clear physical picture still remains the subject of serious
debates.\cite{PRL_076604,PRB_165313} The relationship between the
spin current and the induced spin polarization seems to be a very
subtle issue. The main problem underlying the debates is the
notion of a spin current itself because spin is not a conserved
quantity in spin-orbit coupled systems. Consequently, any approach
that avoids the intricate identification of a more or less
suitable spin current is superior. Such an alternative approach
not only introduces different calculational techniques, but also
suggests alternative interpretations of the effects under
consideration. As widely anticipated, a complete physical
description of spin-related phenomena is provided by microscopic
models based on the spin-density matrix or Keldysh Green functions
together with an analysis of its long-wavelength and low-frequency
limit. These approaches are more general and free from artefact
associated with ambiguous definitions of the spin current. In
addition, spin-charge coupled kinetic equations allow the
treatment of such interesting phenomena as propagating spin
excitations or the relationship between the intrinsic SHE and the
zitterbewegung.

In this report, we propose an alternative approach to the SHE by
deriving spin-charge coupled drift-diffusion equations for a
two-dimensional hole gas (2DHG), which refers to the populated
heavy-hole band of thin p-type quantum wells. The related
heavy-hole Hamiltonian of the cubic Rashba model has the second
quantized form
\begin{eqnarray}
&&H=\sum_{{\bm{k}},\lambda }a_{\bm{k}\lambda }^{\dag}\left[ \varepsilon_{%
\bm{k}}-\varepsilon _{F}\right] a_{%
\bm{k}\lambda }-\sum_{\bm{k},\lambda ,\lambda ^{\prime }}\left(
\hbar{\bm{\omega}}_{
\bm{k}} \cdot {\bm{\sigma }}_{\lambda \lambda ^{\prime }}\right) a_{\bm{k}%
\lambda }^{\dag}a_{\bm{k}\lambda ^{\prime }}\nonumber\\
&&+ u\sum\limits_{{\bm{k}},{\bm{k}}^{\prime}}
\sum\limits_{\lambda}a_{{\bm{k}}\lambda}^{\dag}a_{{\bm{k}}^{\prime}\lambda}
-ie{\bm{E}}\sum\limits_{{\bm{k}},\lambda}\nabla_{\bm{\kappa}}
a^{\dag}_{{\bm{k}}-{\bm{\kappa}}/2\lambda}a_{{\bm{k}}+{\bm{\kappa}}/2\lambda}\mid_{\bm{\kappa}=0},
\label{Hamil}
\end{eqnarray}
where $a_{\bm{k}\lambda}^{\dag}$ ($a_{\bm{k}\lambda}$) denote the
creation (annihilation) operators with in-plane quasi-momentum
$\bm{k}=(k_x,k_y,0)$ and spin $\lambda$. The electric field
${\bm{E}}$ is oriented along the $x$ axis. Furthermore,
$\varepsilon_F$ denotes the Fermi energy, $\bm{\sigma}$ the vector
of Pauli matrices, $\varepsilon_{\bm{k}}=\hbar^2k^2/(2m)$, and $u$
the strength of the 'white-noise' elastic impurity scattering,
which gives rise to the momentum relaxation time $\tau$. Contrary
to a phenomenological approach, we treat elastic scattering on a
full microscopic scope. The spin-orbit coupling is given by
\begin{equation}
\hbar {\bm{\omega}}_{\bm{k}}=\frac{\alpha}{2}
\left[i(k_{+}^3-k_{-}^3),(k_{+}^3+k_{-}^3),0 \right],\label{omega}
\end{equation}
where $k_{\pm}=k_x\pm ik_y$ and $\hbar\omega_k=\alpha k^3$. Based
on the Born approximation for elastic impurity scattering, the
Laplace-transformed kinetic equations for the physical components
of the spin-density matrix have the form~\cite{PRB_165313}
\begin{equation}
sf-\frac{i\hbar}{m}(\bm{\kappa}\cdot\bm{k})f+i{\bm{\omega}}_{\bm{\kappa}}
({\bm{k}})\cdot{\bm{f}}
+\frac{e{\bm{E}}}{\hbar}\nabla_{\bm{k}}f=\frac{1}{\tau}(\overline{f}-f)+f_0,
\label{kin1}
\end{equation}
\begin{eqnarray}
&& s{\bm{f}}+2({\bm{\omega}}_{\bm{k}}\times{\bm{f}})
-\frac{i\hbar}{m}(\bm{\kappa}\cdot\bm{k}){\bm{f}}
+i{\bm{\omega}}_{\bm{\kappa}}
({\bm{k}})f+\frac{e{\bm{E}}}{\hbar}\nabla_{\bm{k}}
{\bm{f}}\nonumber\\
&&=\frac{1}{\tau}(\overline{{\bm{f}}}-{\bm{f}})+\frac{1}{\tau}
\frac{\partial}{\partial\varepsilon_{\bm{k}}}
\overline{f\hbar{\bm{\omega}}_{\bm{k}}}-
\frac{\hbar{\bm{\omega}}_{\bm{k}}}{\tau}
\frac{\partial}{\partial\varepsilon_{\bm{k}}} \overline{f}
+{\bm{f}}_0, \label{kin2}
\end{eqnarray}
where an additional frequency appears
\begin{eqnarray}
{\bm{\omega}}_{\bm{\kappa}}({\bm{k}})&=&\frac{3\alpha}{\hbar}
\left[(k_y^2-k_x^2)\kappa_y-2k_xk_y\kappa_x,\right. \nonumber\\
&&\left. (k_x^2-k_y^2)\kappa_x-2k_xk_y\kappa_y,0 \right],
\end{eqnarray}
which depends on ${\bm{\kappa}}$. The cross line over
${\bm{k}}$-dependent quantities denotes an integration over the
polar angle $\varphi$ of the in-plane vector ${\bm{k}}$. $f_0$ and
${\bm{f}}_0$ denote the initial charge and spin density
components, respectively. The quantum Boltzmann equations
(\ref{kin1}) and (\ref{kin2}) are treated in the long-wavelength
limit in order to derive spin-charge coupled drift-diffusion
equations. To this end, the kinetic equations are written in a
matrix form $A\widehat{f}+{\cal
E}\widehat{f}=B\overline{\widehat{f}}+ \widehat{\delta}$, where
the matrix $A$ collects all contributions that are independent of
the electric field ${\bm{E}}$ and not integrated over the angle
$\varphi$ [$\widehat{f}$ denotes the four component vector
$(f,{\bm{f}})$]. To calculate the matrix $B$ on the right-hand
side of this equation, we assume $\alpha k_F^3\ll
\varepsilon_{k_F}$ and restrict $\kappa$ contributions up to
$\kappa^2$. The matrix equation is solved iteratively in the case
of weak electric field contributions ${\cal E}$ (the matrix ${\cal
E}$ contains first-order derivatives and $\widehat{f}$ is
decomposed according to $\widehat{f}_0+\widehat{f}_1$ with
$\widehat{f}_0\sim E^0$ and $\widehat{f}_1\sim E$). The solution
of the equation is written in the form
\begin{equation}
(1+\overline{C}_1^{-1}\overline{C}_2)^{-1}\overline{C}_1^{-1}
\overline{\widehat{f}}=\widehat{\delta},
\end{equation}
where $\overline{\widehat{f}}_0=\overline{C}_1\widehat{\delta}$
and $\overline{\widehat{f}}_1=\overline{C}_2\widehat{\delta}$. The
general expressions for the transport coefficients are very
cumbersome but simplify considerably in the low-field case and
under the condition $\alpha k_F^3\ll \varepsilon_{k_F}$.

As we are mainly interested in the SHE, let us focus on the
coupling between the spin and charge degrees of freedom. By
applying the outlined schema, we obtain our main analytical result
\begin{equation}
(s+i\frac{v_d}{\sigma_0}\kappa_x+D_0\kappa^2)\overline{f}+i\Gamma_z\kappa_y\overline{f}_z=f_0,\label{e1}
\end{equation}
\begin{equation}
(s+\frac{1}{\tau_{sz}}+i\frac{v_d}{\sigma_0}\kappa_x+D_z\kappa^2)\overline{f}_z
+i\Gamma_0\kappa_y\overline{f}=0,\label{e2}
\end{equation}
with the transport coefficients
\begin{equation}
D_0=\frac{D}{\sigma_0^2},\quad
D_z=D\frac{\sigma_0^2-12\Omega^2}{(\sigma_0^2+4\Omega^2)^2},\label{e10}
\end{equation}
\begin{equation}
\Gamma_0=v_d\frac{9\Omega^2}{2\gamma}\frac{3\sigma_0^2-4\Omega^2}{(\sigma_0^2+4\Omega^2)^2},
\quad \frac{1}{\tau_{sz}}=\frac{4\Omega^2}{\sigma_0\tau},
\end{equation}
\begin{equation}
\Gamma_z=v_d\frac{9\Omega^2}{2\gamma}
\sigma_0^2\frac{4\sigma_0\Omega^2+8\Omega^2-3\sigma_0^2s\tau}
{(\sigma_0s\tau +4\Omega^2)(\sigma_0^2+4\Omega^2)}.
\end{equation}
The parameters introduced in these equations are given by:
$v_d=eE\tau/m$, $\sigma_0=s\tau+1$, $D=v^2\tau/2$,
$\gamma=\varepsilon_k\tau/\hbar$, and $\Omega=\omega_k\tau=K(k)l$
(with $K(k)=\alpha mk^2/\hbar^2$ and the mean-free path
$l=v\tau$). As the contributions $\sim \kappa_x$ do not affect our
analysis, we took them to lowest order in $\Omega$. The
Eqs.~(\ref{e1}) and (\ref{e2}) have been derived for
$\Omega/\gamma\ll 1$ but unrestricted values $\Omega$. In the
absence of the electric field ($E=0$), the Eqs.~(\ref{e1}) and
(\ref{e2}) completely decouple. This decoupling, which applies to
all components of the spin-density matrix, represents a speciality
of the cubic Rashba model.~\cite{PRB_193316,PRB_195330} The
situation is completely different for electrons (linear Rashba
model), for which at zero external fields the charge density
$\overline{f}$ couples to the transverse spin component
$\overline{f}_r=i({\bm{\kappa}}\times{\overline{\bm{f}}})_z$,
whereas $\overline{f}_z$ is connected with the longitudinal
component $\overline{f}_d=i{\bm{\kappa}}\cdot{\overline{\bm{f}}}$.
However, both for the linear and cubic Rashba model additional
couplings arise due to an applied electric field. The related
magnetization gives rise to a magnetoelectric effect that has been
thoroughly investigated in the literature for semiconductors with
spin-orbit interaction.

The time dependence of the coupled spin-charge transport is
calculated by an inverse Laplace transformation of the solution of
Eqs.~(\ref{e1}) and (\ref{e2}). Due to the complicated $s$
dependence of all transport coefficients, a non-Markovian temporal
evolution is expected. However, the straightforward determination
of this complicated time dependence of charge and spin densities
becomes problematic. Eqs.~(\ref{e1}) and (\ref{e2}) are only
solvable by inverse Fourier- and Laplace transformations under
appropriate additional approximations concerning the $s$
dependence. This delicate mathematical problem will be accounted
for in a forthcoming paper. Here, we focus on steady-state
solutions ($\sigma_0=1$).

The character of the coupled spin-charge transport strongly
depends on the strength of the spin-orbit coupling, which is
expressed by the dimensionless parameter $\Omega=K(k)l$. It is the
most striking feature of the drift-diffusion Eqs.~(\ref{e1}) and
(\ref{e2}) that the character of the spin transport changes
radically with increasing coupling strength $\Omega$. The
appearance of such a crossover is due to the unusual expression
for the diffusion coefficient $D_z$ in Eq.~(\ref{e10}), which has
recently been derived by an alternative
approach.~\cite{PRB_193316} Moreover, the very same result is also
obtained for the linear Rashba model, when the frequency
$\omega_k$ is appropriately redefined. With increasing spin-orbit
coupling $K(k)$ or relaxation time $\tau$, the diffusion
coefficient $D_z$ changes its sign. A negative diffusion
coefficient indicates an instability of the spin system. Under
this condition, spin diffusion has the tendency to strengthen
initial spin fluctuations. The competition between this self
strengthening and spin relaxation processes results in a spatial
oscillatory spin pattern. Going from weak ($\Omega<1/\sqrt{12}$)
to strong ($\Omega>1/\sqrt{12}$) spin-orbit coupling, we observe a
transition in the spin system from a diffusive behavior to a
ballistic regime. We shall show that ballistic spin transport is
characterized by undamped spin oscillations.

To be specific, let us solve Eqs.~(\ref{e1}) and (\ref{e2}) for a
stripe geometry ($-L_0\leq y\leq L_0$). The inverse Fourier
transformation is accomplished by the replacement
$\kappa_y\rightarrow i\partial/\partial y$, whereas along the $x$
axis all quantities are constant. The resulting set of
differential equations is easily solved. For the boundary
condition $\overline{f}(-L_0)=\overline{f}(L_0)=f_0$ and
$\overline{f}_z(-L_0)=\overline{f}_z(L_0)=0$, we obtain
\begin{equation}
\overline{f}_z(k,y)=f_0\frac{\tau_{sz}\Gamma_0}{D_0}\,\frac{y\sinh(L_0/l_0)-L_0\sinh(y/l_0)}
{\sinh(L_0/l_0)},\label{e13}
\end{equation}
where $l_0=\sqrt{D_z\tau_{sz}}$.
\begin{figure}
\centerline{\epsfig{file=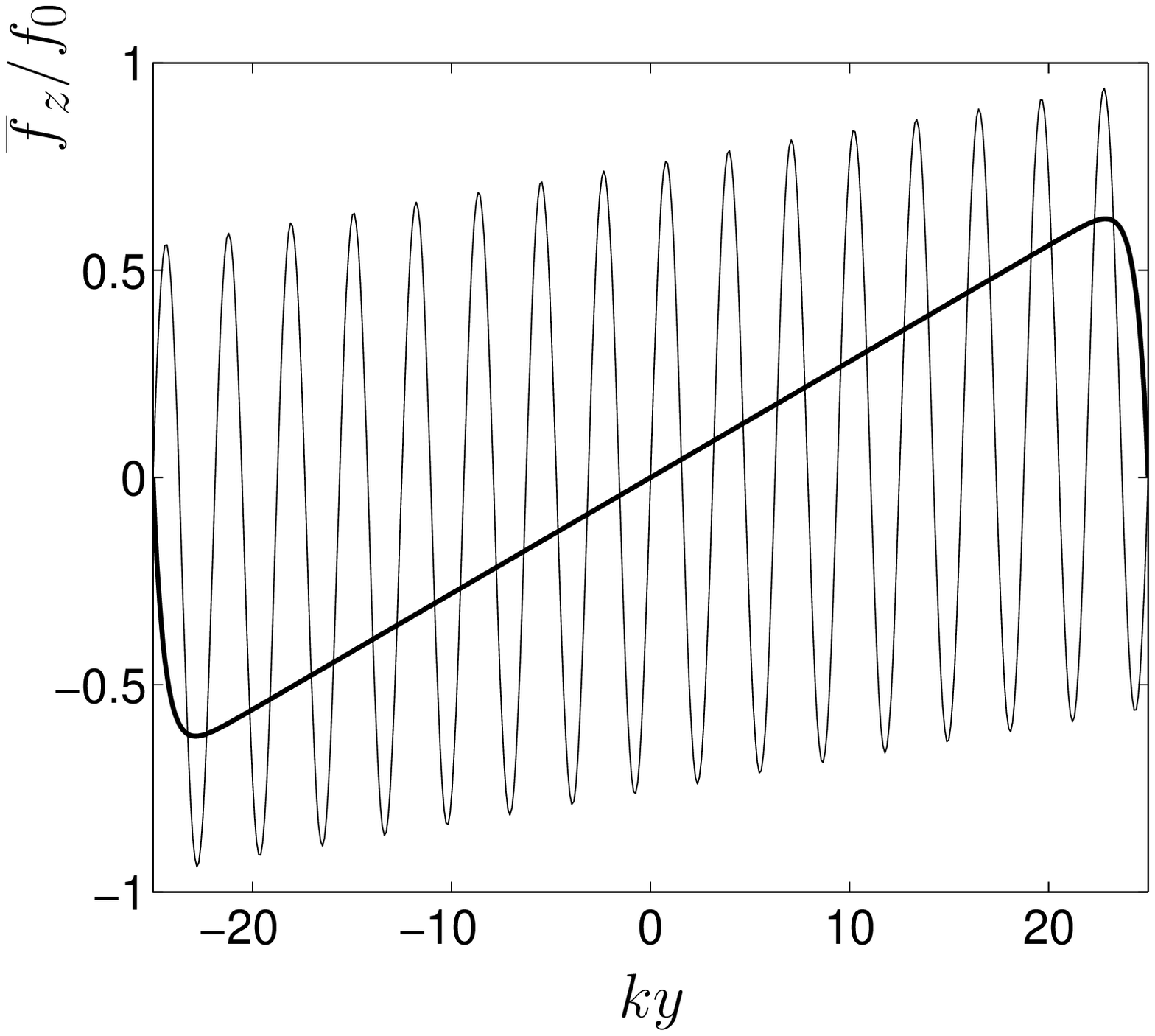,width=7.0cm}} \caption{Out-of
plane spin polarization induced by an electric field applied
parallel to the stripe of a 2DHG (with $\gamma=0.5$). The thick
and thin lines were calculated with $\Omega=0.25$ and
$\Omega=0.5$, respectively.} \label{fig1}
\end{figure}
An interesting effect, which we do not follow up in this paper,
results from the $y$ dependence of the charge density
$\overline{f}$ that is strongly affected by the boundary condition
and that gives rise to a self-consistent electric field oriented
along the $y$ direction. In Fig.~1, the thick line illustrates the
result for the spin polarization in the diffusive regime
($D_z>0$), when the spin-orbit coupling is weak
$\Omega<1/\sqrt{12}$. The electric field aligned along the $x$
axis induces a spin polarization at the edges of the stripe. This
SHE has received a great deal of attention. Most popular is the
description by means of a spin current oriented perpendicular to
the electric field. Many theoretical studies (see, for instance,
Ref.[\onlinecite{PRB_085308,PRB_193316}]) introduced the spin
current $\widehat{J}_{\mu}^{i}$ by a symmetrized product of spin
and velocity operators
$(\widehat{v}_{\mu}\sigma^{i}+\sigma^{i}\widehat{v}_{\mu})/2$. It
was claimed that at least for the cubic Rashba model, the SHE
introduced in such a manner is robust against disorder.
Experimental results \cite{PRL_047204} seem to confirm this
physical picture. However, there is a principal difficulty with
such an approach. The above mention definition of the current is
not sufficiently general. It would completely fail for any hopping
transport problem, for which the eigenstates have no dispersion.
This definition only applies, whenever the Hamiltonian commutes
with the dipole operator. Obviously, this is not the case for the
Rashba Hamiltonian. Consequently, it is necessary to go back to
the more general definition, which expresses the current by the
time derivative of the dipole
operator.~\cite{PRL_076604,PRB_165313} This more general treatment
of the spin current reveals a close relationship between the
field-induced spin accumulation and the spin current expressed by
a quasi-chemical potential. From a technical point of view, the
current that applies in its most general form to a homogeneous
system is calculated from the quantity
$\nabla_{\bm{\kappa}}\widehat{f}({\bm{k}},{\bm{\kappa}})$ at
${\bm{\kappa}}={\bm{0}}$ and not only from the density matrix
$\widehat{f}({\bm{k}})$. Based on this general definition of the
spin current, it was concluded that an electric-field-induced
steady-state spin-Hall current does not exist in the cubic Rashba
model.~\cite{JPCM_7497} On the other hand, a SHE was demonstrated
by a recent experiment.~\cite{PRL_047204} This calamity indicates
that the notion of a spin current is not useful for studying the
SHE. An alternative, which is proposed in this paper, determines
the spin accumulation from quantum kinetic equations or the
associated spin-charge coupled drift-diffusion equations. The
specific difficulty of the latter approach is the formulation of
appropriate boundary conditions.~\cite{PRB_113309,PRB_115331}

The character of the SHE dramatically changes in strongly
spin-orbit coupled systems ($\Omega>1/\sqrt{12}$ so that $D_z<0$).
The result is illustrated in Fig.~1 by the thin line. A
spin-coherent standing wave travels through the stripe. It is
remarkable that these oscillations are not damped although a
finite elastic scattering is present. The occurrence of a periodic
spin pattern is not unusual and has been investigated in the
literature.~\cite{PRB_155317,PRB_033316,PRB_195308,PRB_205307} The
novelty here is that such an oscillatory spin pattern can be
induced by an electric field. The rapid variation of the out-of
plane spin polarization induces a magnetic field that leads to
circulating microscopic currents. The retroaction of these
currents on spin may result in a finite damping of spin
oscillations.

In the strong-coupling limit ($\Omega=K(k)l\gg 1$), the
oscillatory spin pattern changes on a length scale $K^{-1}(k)$
that is much smaller than the mean-free path $l$. This fact
conflicts with basic assumptions of the drift-diffusion approach,
which is only applicable for diffusion lengths much smaller than
the mean-free path. Although macroscopic transport equations were
found to be valid even when the spin-diffusion length is somewhat
less than the mean-free path \cite{PRB_212410}, it is
indispensable to treat this point in detail. Large values for
$\Omega$ give rise to spin-relaxation times $\tau_{sz}$, which are
much smaller than the elastic scattering time $\tau$. This
condition characterizes the ballistic spin regime. Therefore, we
go back to the kinetic Eqs.~(\ref{kin1}) and (\ref{kin2}) and
solve them under the condition $\tau\rightarrow \infty$ and to
first order in the electric field. For the out-of plane spin
polarization, we obtain
\begin{equation}
(\sigma^2+4\omega_k^2)f_z+\frac{eE\sigma}{\hbar}\frac{\partial}{\partial
k_x}f_z=\frac{2i}{\sigma}({\bm{\omega}}_{{\bm{k}}x}{\bm{\omega}}_{{\bm{\kappa}}y}
-{\bm{\omega}}_{{\bm{k}}y}{\bm{\omega}}_{{\bm{\kappa}}x})f_0,
\end{equation}
where $\sigma=s-i{\bm{\kappa}}\cdot{\bm{v}}$. Calculating the
inverse Laplace and Fourier transformation and integrating over
the angle $\varphi$, we arrive at the analytical solution
\begin{eqnarray}
&&f_z(k,{\bm{r}},t)=\frac{3f_0}{16\pi r}\frac{\partial}{\partial
k}\left[\frac{eE}{\varepsilon_k}\left(\frac{\sin 2K(k)r}{2K(k)r}
-\cos 2K(k)r \right) \right]\nonumber\\
&& \times\left\{
\begin{array}{ccc}
  \delta(y-\sqrt{(vt)^2-x^2}) & {\rm for} & 0<y\leq vt \\
  -\delta(y+\sqrt{(vt)^2-x^2}) & {\rm for} & vt\leq y<0 \\
\end{array}\right.\,,
\end{eqnarray}
which describes the field-induced spin polarization that occurs,
when initially a drop of charge carriers is injected into the 2DHG
at the position ${\bm{r}}={\bm{0}}$.
\begin{figure}
\centerline{\epsfig{file=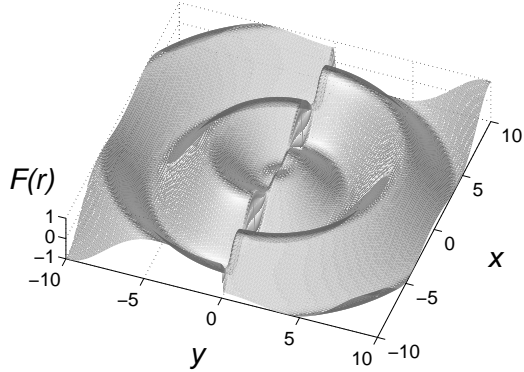,width=7.0cm}} \caption{The
amplitude $F(r)=\eta x{\rm d}((\sin x/x-\cos x)/x)/{\rm d}x$ in
dependence on $r$ with $x=2K(k_F)r$ and $K(k_F)=0.5$. For the
factor $\eta$, we have: $\eta = \pm 1$ for $y \gtrless 0$.}
\label{fig2}
\end{figure}
A $\delta$-like wave front of spin polarization travels through
the homogeneous 2DHG. The amplitude of this narrow wave front
oscillates as illustrated in Fig.~2. As expected, the relief is
antisymmetric with respect to the $y$ axis. Most interesting for
our comparison with the above drift-diffusion approach is the
observation that the wavelength of the spatial spin pattern is of
the order of $K^{-1}(k_F)$. Therefore, this exact analytical
result confirms the existence of a field-induced long-living spin
pattern in strongly spin-orbit coupled systems as predicted by the
drift-diffusion equations.

Our study of electric-field induced spin phenomena revealed a
close relationship between the SHE and spin-coherent oscillations.
We compare this conclusion with recent results that demonstrated
that the intrinsic SHE and the zitterbewegung are essentially the
same kind of phenomena.~\cite{PRB_205307} Consequently, the
question arises whether the above treated spin-coherent waves have
to be identified with the zitterbewegung, which is a purely
relativistic effect. Indeed, both kinds of oscillatory spin
excitations exhibit common features. The characteristic wavelength
of both types of oscillations amounts about $100$~nm (calculated
by adopting the typical values $\alpha m/\hbar^{2}\sim 2$~nm and
$k_F\sim 0.1$~nm). Moreover, both the zitterbewegung
\cite{PRL_206801} and the spin-coherent waves [cf.
Eq.~(\ref{e13})] are resonantly enhanced, whenever the width of
the stripe matches a characteristic wavelength of the spin
excitation. However, some features of spin-coherent waves are not
compatible with such an identification with the zitterbewegung.
The dispersion relation of coupled spin-charge excitations is
calculated from the vanishing determinant of Eqs.~(\ref{e1}) and
(\ref{e2}). In general, one obtains not only spin-coherent
solutions but also damped excitations, whereas the transition
between them could be driven by the electric field. In addition,
the spin-coherent waves that appear at strong spin-orbit coupling
are separated from the SHE by a sharp threshold. For the
zitterbewegung such a threshold is not expected as its existence
is solely due to at least two energy bands separated by a nonzero
gap. We think that the interesting study of the relationship
between spin-coherent waves, the zitterbewegung, and long-living
spin-coherent states \cite{PRB_155317} will continue in the near
future.

The experimental observation of the field-induced spin-coherent
waves should be possible by high-resolution scanning-probe
microscopy imaging techniques. The direct experimental proof of
this effect would fascilitate developments both in spintronics and
basic research.

This work was supported by the Deutsche Forschungsgemeinschaft and
the Russian Foundation of Basic Research under the grant number
05-02-04004.


\end{document}